\documentclass[aps,prl,reprint,superscriptaddress,floatfix]{revtex4-2}
\usepackage[colorlinks=true,allcolors=blue]{hyperref} 
\usepackage{graphicx} 
\usepackage{bm}
\usepackage{amsmath, amssymb}
\usepackage{braket}
\usepackage{appendix}
\usepackage{xcolor}
\usepackage{siunitx}

\newcommand{\Figref}[1]{Fig.~\ref{#1}}
\newcommand{\Figsref}[1]{Figs.~\ref{#1}}
\newcommand{\Eqref}[1]{Eq.~(\ref{#1})}

\newcommand{\lb}{\left(}
\newcommand{\rb}{\right)}

\begin{document}
\title{Rotation-to-translation conversion by geometric asymmetry in viscoelastic fluids}

\author{Takuya Kobayashi}\email{tkobaya@stanford.edu}
\affiliation{Department of Chemical Engineering, Kyoto University, Kyoto 615-8510, Japan}
\affiliation{Department of Chemical Engineering, Stanford University, Stanford, CA 94305, USA}

\author{Hiroki Kitano}
\affiliation{Department of Chemical Engineering, Kyoto University, Kyoto 615-8510, Japan}

\author{Ryoichi Yamamoto}\email{ryoichi@cheme.kyoto-u.ac.jp}
\affiliation{Department of Chemical Engineering, Kyoto University, Kyoto 615-8510, Japan}

\date{\today}

\begin{abstract}
Microscale locomotion in Newtonian fluids is constrained by kinematic reversibility. Here we show that viscoelasticity provides a distinct route: an achiral fore–aft asymmetric body rotating in a viscoelastic fluid generates net translation through normal-stress-driven secondary flows. Direct numerical simulations combined with scaling analysis reveal the universal scaling law $V\sim{\rm Wi}\cdot S$, where $\rm Wi$ is the Weissenberg number and $S$ is the skewness of the axial volume distribution. This result identifies a minimal geometric principle for rotation-induced propulsion in viscoelastic fluids, and suggests a route for active microrheology via propulsion-speed measurements.

\end{abstract}

\maketitle

\textit{Introduction}---Microscale propulsion is governed by symmetry-imposed constraints. In Newtonian fluids, kinematic reversibility forbids locomotion by reciprocal shape deformations, as formalized by Purcell's scallop theorem~\cite{Purcell1977-ab}. Consequently, propulsion requires breaking time-reversal symmetry, typically through non-reciprocal actuation or geometric chirality, such as rotating helical filaments~\cite{Lauga2009-jt, Lauga2011-kp}. 
The constraints fundamentally limit the design of microswimmers and microrobots, making symmetry breaking a central requirement for locomotion at low Reynolds numbers and motivating the development of alternative propulsion mechanisms.

Microswimmers have been mainly studied in Newtonian fluids, whereas propulsion in non-Newtonian fluids remains poorly understood, even though most biological fluids exhibit non-Newtonian rheology~\cite{Li2021-fv}, with important implications for biological transport and related applications such as targeted drug delivery~\cite{Patra2013-lm, Gao2014-xt}.
In such fluids, the scallop theorem breaks down: non-Newtonian stresses break time-reversal symmetry, thereby violating kinematic reversibility and enabling propulsion even under reciprocal actuation~\cite{Lauga2009-df, Keim2012-dd, Qiu2014-nm, Datt2018-pe, Han2020-hy, Rogowski2021-no, Kobayashi2024-ke}. 

\begin{figure}[b]
    \centering
    \includegraphics[width=\linewidth]{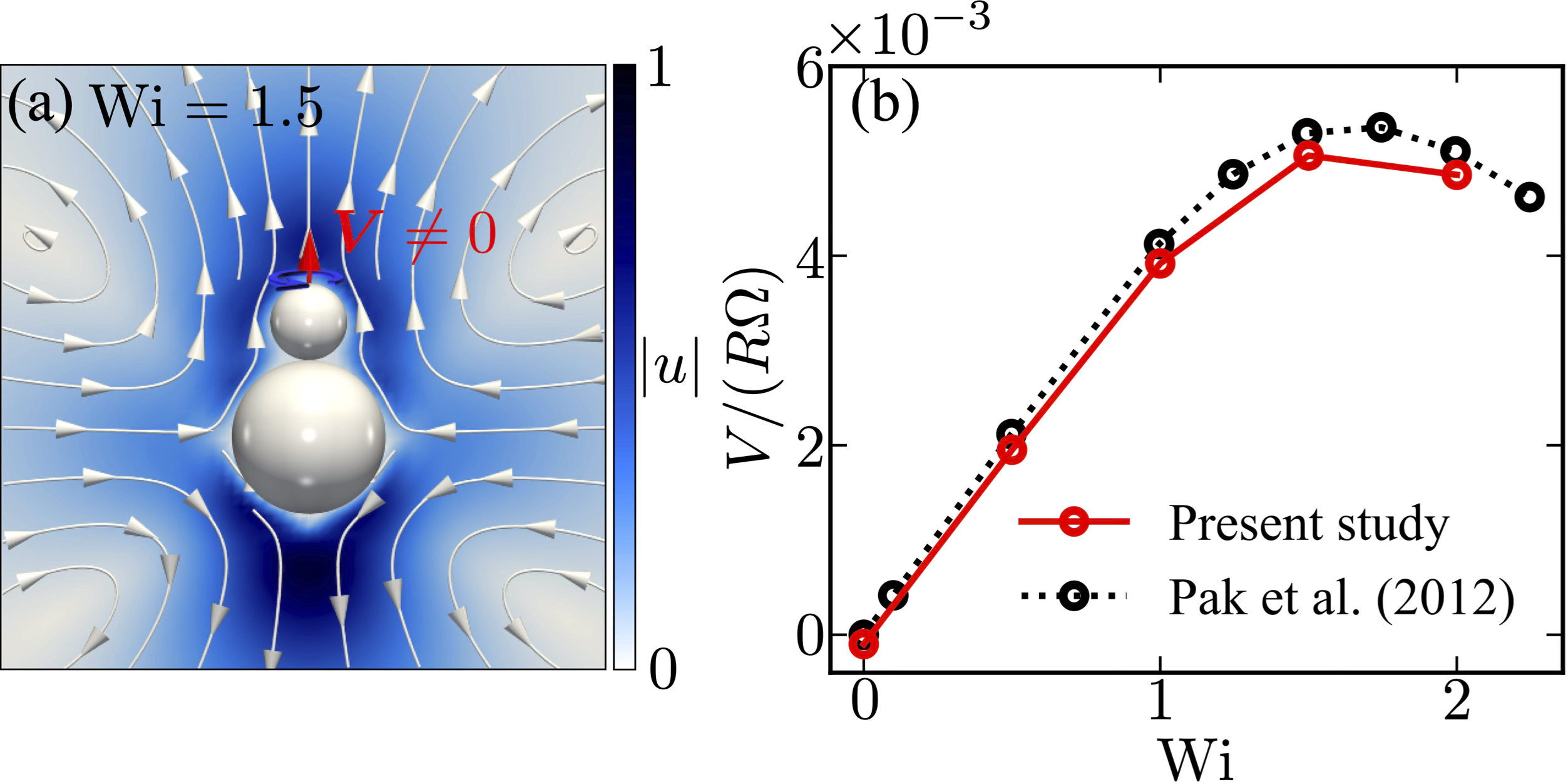}
    \caption{(a) Flow field around a rotating snowman-shaped particle. 
    The velocity field is projected onto the meridional plane containing the rotation axis, and the plotted magnitude excludes the azimuthal component.
    In viscoelastic fluids, elastic normal stresses generate a secondary flow, analogous to the Weissenberg effect. 
    This flow breaks fore--aft symmetry and converts pure rotation into net translation for fore--aft asymmetric bodies.
    (b) Dimensionless propulsion speed $V/(R\Omega)$ as a function of ${\rm Wi}$, showing quantitative agreement with Ref.~\cite{Pak2012-rk}. The particle Reynolds number is ${\rm Re}=0.1$.
    }
    \label{fig1}
\end{figure}

Rotational motion is particularly appealing as it represents the simplest form of actuation, involving a single degree of freedom. 
In Newtonian fluids, rotation can produce translation only when geometric chirality, as in a screw, couples the two motions; however, reciprocal back-and-forth rotation produces no net displacement.
Viscoelastic fluids provide an alternative route:
fore–aft asymmetric particles (e.g., snowman-like shapes; see \Figref{fig1}(a)) can convert rotation into translation~\cite{Pak2012-rk, Puente-Velazquez2019-fs, Rogowski2021-no, Binagia2021-fi, Kroo2022-qm}. 
Rotation generates normal stress differences that drive secondary flows, analogous to the Weissenberg effect~\cite{Weissenberg1947-lx}, which are absent in Newtonian fluids at zero Reynolds number. 
Here, the ``secondary flow'' refers to the radial and axial velocity components in the meridional plane, i.e., the plane containing the rotation axis, in contrast to the primary azimuthal flow directly generated by particle rotation~\cite{Fosdick1980-lt}. For a fore--aft asymmetric body, this elastic secondary flow produces a net translation along the rotation axis.

Figure~\ref{fig1} illustrates this mechanism for a snowman-shaped particle. At ${\rm Wi}=0$, rotation generates only the primary azimuthal flow and no net translation. At finite ${\rm Wi}$, elastic normal stresses drive a secondary flow [\Figref{fig1}(a)]. This flow couples to the fore--aft asymmetry of the particle and converts pure rotation into net translation [\Figref{fig1}(b)].
By the same mechanism, microswimmers can achieve enhanced propulsion in viscoelastic fluids~\cite{Binagia2020-dn, Housiadas2021-qe, Kobayashi2023-ad, Kobayashi2024-tn}.
Despite these advances, a general principle linking geometry to propulsion in complex fluids remains lacking. 

While \citet{Pak2012-rk} derived $V \sim {\rm Wi}$ for snowman bodies via the reciprocal theorem, a universal geometric descriptor linking arbitrary shape to propulsion has remained unknown. 
In this Letter, we demonstrate that propulsion can be achieved without geometric chirality, arising solely from the coupling between fluid elasticity and geometric fore–aft asymmetry. 
By introducing a scalar measure $S$ of geometric asymmetry based on the skewness of the particle shape and systematically analyzing 
28 rigid axisymmetric particles of distinct shapes,
we show that the propulsion speed is uniquely determined by a leading-order scaling $V \sim{\rm Wi}\cdot S$.
This establishes a minimal and predictive framework for rotation-induced propulsion in complex fluids. 
Our results generalize propulsion to arbitrary shapes and provide a unified design principle for microswimmers in viscoelastic fluids, while suggesting new routes for microrheological probing based on rotation-induced transport~\cite{Kroo2022-qm, Ghosh2018-mp, Ghosh2021-cv, Chiu2026-am}.

\textit{Numerical method}---We simulate the fluid-particle hydrodynamic coupling using the Smoothed Profile (SP) method~\cite{Yamamoto2021-oe, Kobayashi2023-ad}. In this method, the sharp fluid-particle boundary is replaced by a continuous diffuse interface of thickness $\xi$, by introducing a smoothed profile function $\phi \in [0, 1]$, which is 1 inside the particle and 0 in the surrounding fluid. The mathematical definition of $\phi$ can be found in Ref.~\cite{Kobayashi2023-ad}.
The total velocity field $\bm{u}=(1-\phi)\bm{u}_f+\phi\bm{u}_p$, composed of the host fluid velocity $(1-\phi)\bm{u}_f$ and the rigid-body particle contribution $\phi\bm{u}_p$, obeys
\begin{gather}
\bm{\nabla}\cdot\bm{u} = 0,\qquad
\rho\lb\frac{\partial}{\partial t}+\bm{u}\cdot\bm{\nabla}\rb\bm{u}
=\bm{\nabla}\cdot\bm{\Sigma}
+\rho\phi\bm{f}_p,
\end{gather}
where $\rho$ is the fluid density, $\bm{\Sigma}$ the stress tensor, and $\phi\bm{f}_p$ a body force enforcing particle rigidity.
For a viscoelastic fluid, we adopt the Oldroyd-B model
\begin{align}
\bm{\Sigma}=-p\bm{1}+\bm{\tau}_s+\bm{\tau}_p ,
\end{align}
where the solvent and polymer stresses satisfy
\begin{align}
\bm{\tau}_s=\eta_s\dot{\bm{\gamma}}, \qquad
\bm{\tau}_p+\lambda \stackrel{\triangledown}{\bm{\tau}}_p=\eta_p\dot{\bm{\gamma}} ,
\end{align}
with $\dot{\bm{\gamma}}=\bm{\nabla}\bm{u}+(\bm{\nabla}\bm{u})^{T}$ the strain-rate tensor, total viscosity $\eta = \eta_s + \eta_p$, $\eta_s$ and $\eta_p$ the solvent and polymer viscosities, the viscosity ratio $\beta \equiv \eta_s / \eta = 0.5$, and $\lambda$ the relaxation time. The upper-convected derivative is defined as
\begin{align}
\stackrel{\triangledown}{\bm{\tau}}
= \lb\frac{\partial }{\partial t}
+\bm{u}\cdot\bm{\nabla}\rb\bm{\tau}
-(\bm{\nabla}\bm{u})^{T}\cdot\bm{\tau}
-\bm{\tau}\cdot(\bm{\nabla}\bm{u}) .
\end{align}

\begin{figure}[tb]
    \centering
    \includegraphics[width=\linewidth]{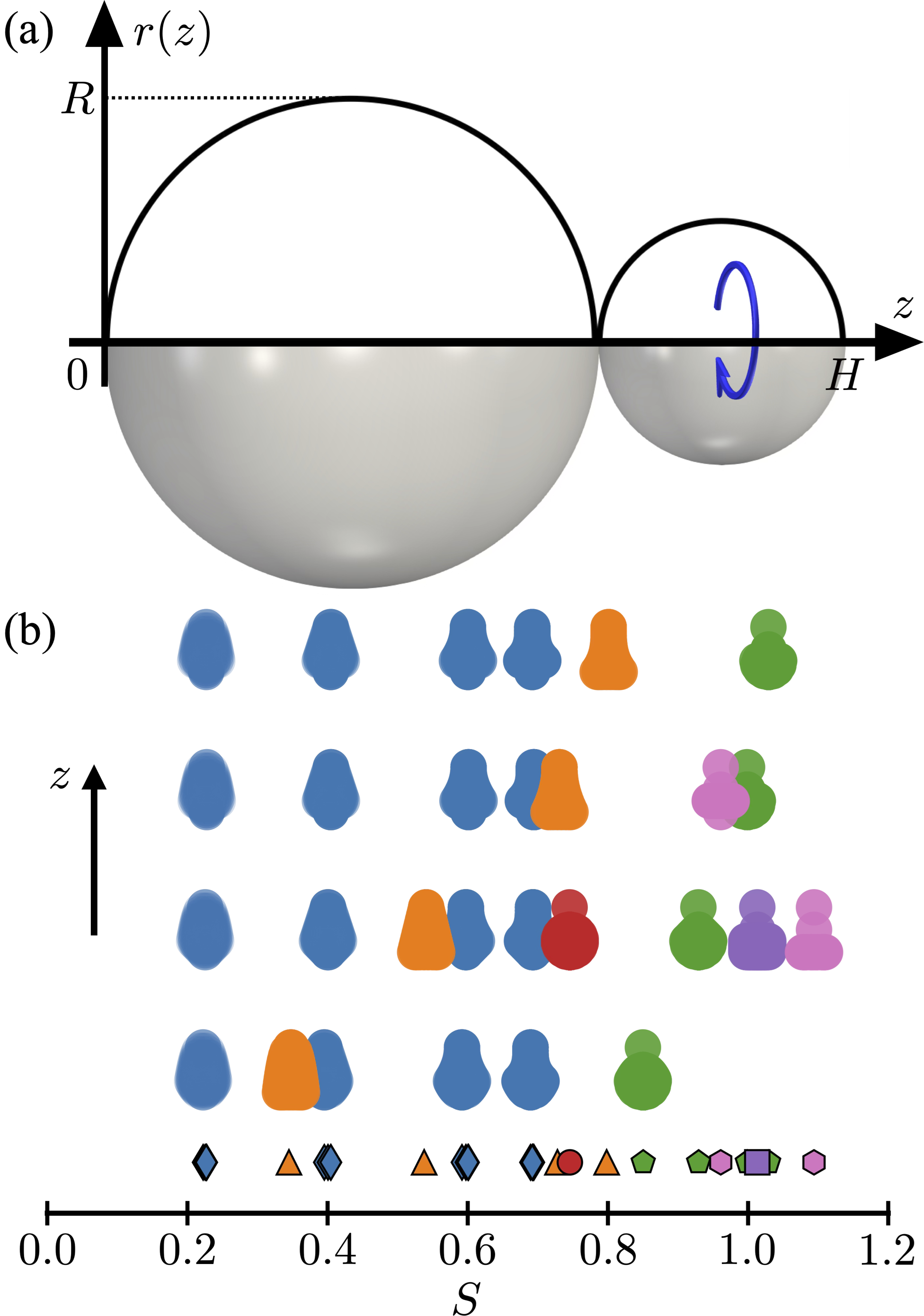}
    \caption{(a) Geometric construction of a fore--aft asymmetric particle, obtained by rotating an axial profile about the symmetry axis $\bm{e}_z$. 
    (b) Representative particle shapes used in the simulations, ordered by the geometric asymmetry parameter $S$. The marker below each shape labels that geometry and is used for the corresponding simulation data points in \Figsref{fig3} and~\ref{fig4} to identify the same geometry.
    }
    \label{fig2}
\end{figure}

\begin{figure}[tb]
    \centering
    \includegraphics[width=\linewidth]{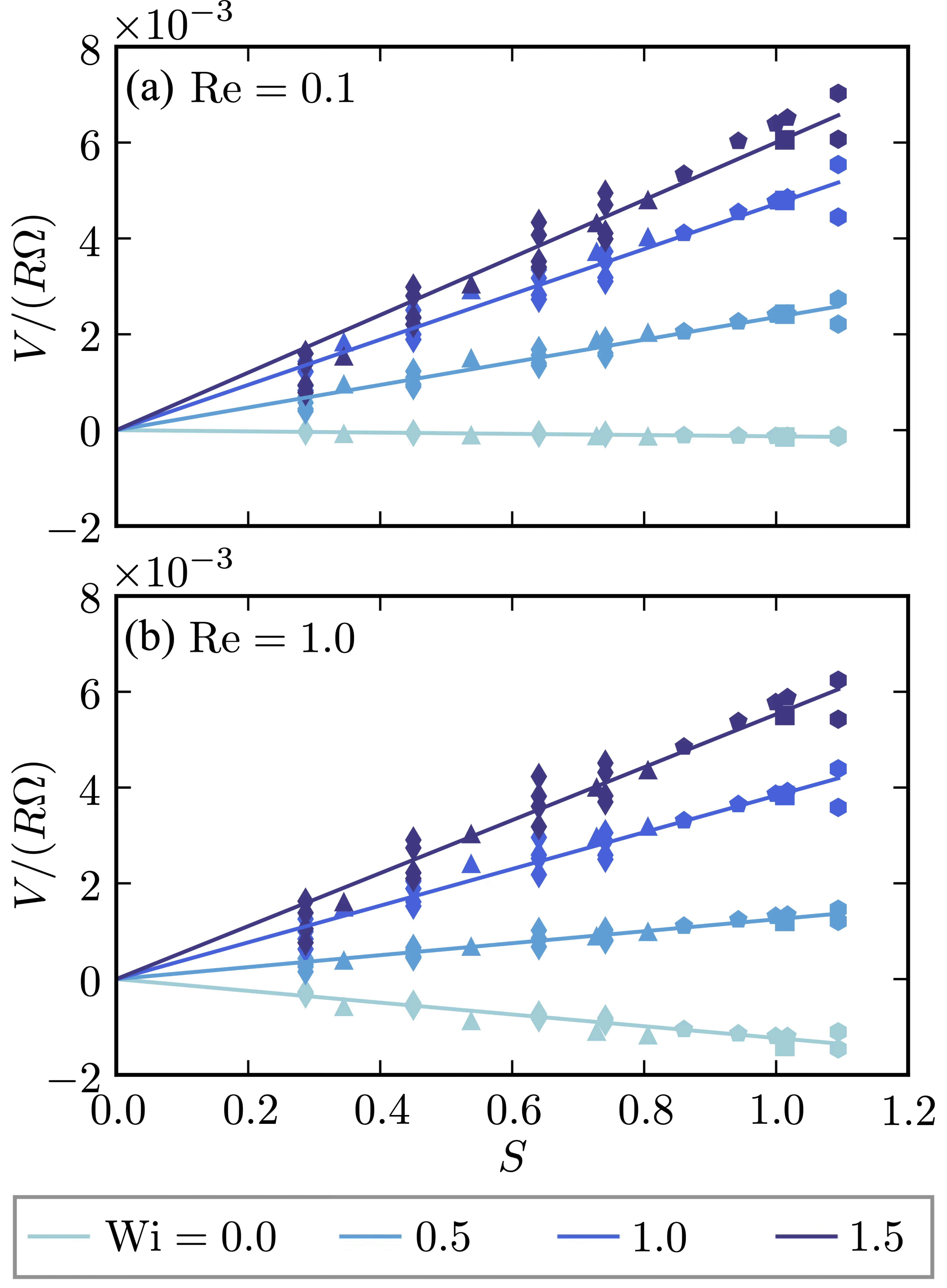}
    \caption{Dimensionless propulsion speed $V/(R\Omega)$ as a function of the geometric asymmetry parameter $S$ for various Weissenberg numbers. Colors denote ${\rm Wi}$, and markers denote the particle shapes defined in Fig.~\ref{fig2}(b).
    The particle Reynolds numbers are (a) $\rm Re = 0.1$ and (b) $\rm Re = 1.0$.
    At $\rm Wi = 0$, the nonzero velocity observed at $\rm Re = 1.0$ originates from inertia-driven secondary flow. Negative values indicate backward propulsion due to inertia-driven flow.
    For ${\rm Wi}>0$, the weak dependence on ${\rm Re}$ indicates that propulsion is dominated by elastic normal-stress-driven secondary flow.
    }
    \label{fig3}
\end{figure}

\begin{figure}[tb]
    \centering
    \includegraphics[width=\linewidth]{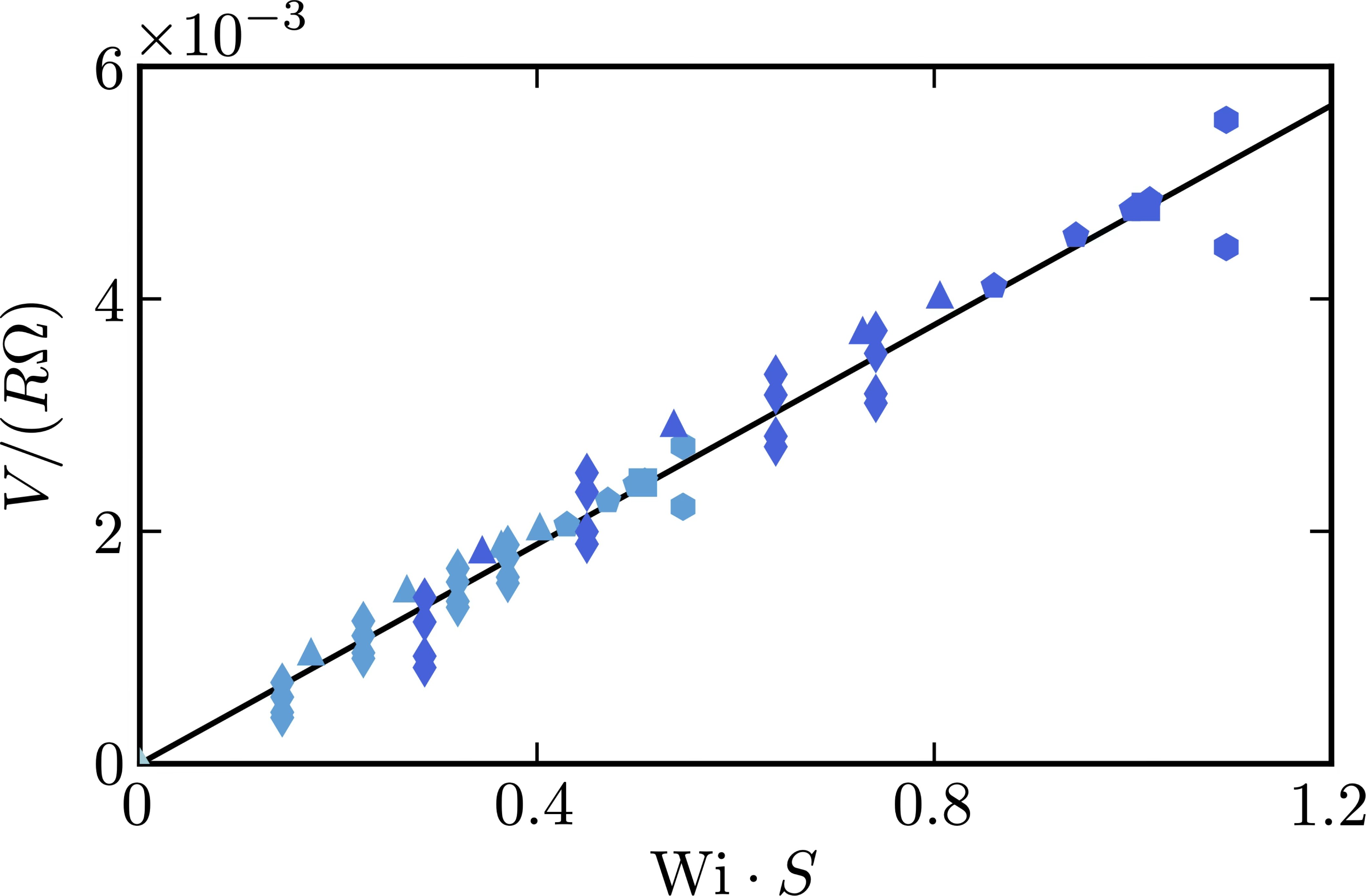}
    \caption{Master curve for the propulsion velocity. 
    Dimensionless propulsion speed $V/(R\Omega)$ as a function of ${\rm Wi}\cdot S$ for all simulated shapes at ${\rm Re}=0.1$ and ${\rm Wi}\leq1$, where the perturbative analysis applies.
    The collapse onto a single linear master curve confirms the scaling $V\sim {\rm Wi}\cdot S$ (\Eqref{eq:scaling}).
    Colors denote ${\rm Wi}$, and markers denote the particle shapes defined in Fig.~\ref{fig2}(b).
    Solid line is a linear fit through the origin.}
    \label{fig4}
\end{figure}

\textit{Geometrical asymmetry}---To quantify the fore–aft asymmetry of particles, we introduce a dimensionless scalar measure based on the axial volume distribution. As shown in Fig.~\ref{fig2}(a), each particle is generated by rotating an axial profile about the symmetry axis ($z$-axis). Figure~\ref{fig2}(b) shows representative shapes used in the simulations. The symbols assigned to these shapes are used consistently in Figs.~\ref{fig3} and~\ref{fig4}, so that the dependence of propulsion on geometry can be tracked across different values of ${\rm Wi}$ and ${\rm Re}$.
For convenience, we normalize $r(z)$ such that
\begin{align*}
    \int_0^H r^2(z) \, dz = 1,
\end{align*}
so that $r^2(z)$ can be interpreted as a probability density along the axis. Note that $r(z)$ is a normalized axial profile (not the geometric radius), and the geometric radius is $\sqrt{A(z)/\pi}$ with $A(z) \propto r^2(z)$.

The asymmetry is quantified by a weighted integral of the form
\begin{align}\label{eq:S}
    S = \frac{1}{\sigma^3} \int_0^H (z - \mu)^3 r^2(z) \, dz,
\end{align}
where
\begin{align*}
    \mu = \int_0^H z r^2(z) \, dz, \quad
    \sigma^2 = \int_0^H (z - \mu)^2 r^2(z) \, dz
\end{align*}
denote the centroid and variance of the axial volume distribution, respectively.
The parameter $S$ corresponds to the skewness (i.e., dimensionless third central moment) of this distribution: $S = 0$ for fore--aft symmetric shapes and increases with asymmetry [see \Figref{fig2}(b)]. As the lowest-order moment of the volume distribution, $S$ provides a minimal and scale-independent characterization of geometric asymmetry.
By construction, $S$ vanishes for fore--aft symmetric profiles and changes sign under fore--aft reflection.

\textit{Linear scaling of propulsion with $S$}---We perform simulations for a single fore--aft asymmetric particle with maximum radius $R=8\Delta$ ($\Delta$ represents the grid spacing) and length $H = 24\Delta$ inside a cubic simulation box of length $L = 128\Delta$, with periodic boundary conditions in all directions. The particle Reynolds number and Weissenberg number are defined as ${\rm Re} \equiv \rho R^2\Omega/\eta$ and ${\rm Wi}\equiv \lambda \Omega$, respectively. 
Our results quantitatively agree with previous studies in the Stokes limit ($\rm Re = 0$) for an unbounded fluid~\cite{Pak2012-rk}, confirming that both finite-size effects and inertia ($\rm Re = 0.1$) are negligible in the present simulations [see \Figref{fig1}(b)].
In all simulations, the particle was driven by a prescribed angular velocity $\Omega$ about its symmetry axis and was allowed to translate freely along that axis under the force-free condition. The propulsion speed was measured from the steady translational velocity.
We examine whether the geometric dependence observed for a snowman particle extends to arbitrary fore--aft asymmetric shapes~\cite{Pak2012-rk}. We simulated 28 rigid axisymmetric particles with different axial volume distributions but the same maximum radius $R$ and length $H$. 
We varied ${\rm Wi}\in[0,1.5]$ and ${\rm Re}\in[0.1,1.0]$.

Figure~\ref{fig3} shows the dimensionless propulsion speed $V/(R\Omega)$ as a function of $S$ for several values of ${\rm Wi}$. For each fixed ${\rm Wi}$, the propulsion speed increases almost linearly with $S$: more strongly fore--aft asymmetric particles propel faster, whereas nearly symmetric particles exhibit little propulsion.
The DNS results are well described by straight lines passing through the origin, indicating the empirical leading-order relation $V \sim S$ at fixed ${\rm Wi}$.
Thus, the skewness $S$ of the axial volume distribution emerges as the relevant lowest-order geometric descriptor, showing that propulsion is a generic consequence of fore--aft asymmetry rather than a shape-specific effect.

Even in Newtonian fluids, fluid inertia can break the kinematic reversibility~\cite{Lauga2007-va} and allow a rotating fore--aft asymmetric particle to self-propel~\cite{Nadal2014-zl, Shen2025-xr}.
This contribution is distinct from the elastic propulsion studied here. As shown in \Figref{fig3}(b), the nonzero velocity at ${\rm Wi}=0$ and ${\rm Re}=1.0$ originates from inertia-driven secondary flow, whereas the ${\rm Wi}>0$ data depend only weakly on ${\rm Re}$, indicating dominance of elastic normal-stress-driven secondary flow, i.e., the Weissenberg-effect-like mechanism. 
The two mechanisms drive motion in opposite directions: the same geometric asymmetry that propels a particle forward with the larger end at the front in a Newtonian fluid drives it backward in a viscoelastic fluid. Such reversal is consistent with the opposite sense of inertial and elastic secondary flows~\cite{Bird1987-qc}. 

This reversal can be understood from the sign of the effective normal-stress anisotropy generated by rotational motion. In viscoelastic fluids, polymer stresses typically satisfy $\tau_{p, rr} - \tau_{p, \theta\theta} < 0$, producing an inward secondary flow through the hoop-stress mechanism. By contrast, in inertial Newtonian flows, the inertial stress $\bm{\Sigma}^{\text{inertia}} \equiv -\rho\bm{uu}$ is dominated by the azimuthal component $u_\theta > u_r$, yielding the opposite anisotropy $\Sigma^{\text{inertia}}_{rr} - \Sigma^{\text{inertia}}_{\theta\theta} > 0$ and hence an outward secondary flow. Fore--aft asymmetry rectifies these opposite secondary flows into propulsion in opposite directions.
In both cases, the resulting propulsion is governed by the same fore--aft asymmetry parameter $S$, although the underlying physical origin differs.

Importantly, Fig.~\ref{fig3} also shows that the slope of $V$ versus $S$ increases with ${\rm Wi}$. The empirical relation $V\sim S$ therefore identifies the relevant geometric descriptor but does not yet provide the full scaling law. We next show that a reciprocal theorem analysis predicts this slope to be proportional to ${\rm Wi}$, yielding the central result $V\sim{\rm Wi}\cdot S$.

\textit{Scaling analysis}---The DNS results identify $S$ as the leading geometric descriptor and show that the proportionality coefficient increases with ${\rm Wi}$. We now derive this dependence in the Stokes ($\rm Re = 0$) and low-${\rm Wi}$ limits using the reciprocal theorem~\cite{Pak2012-rk,Masoud2019-pz}. Non-dimensionalizing lengths by $R$, time by $\Omega^{-1}$, and stresses by $\eta\Omega$, the leading-order propulsion velocity is
\begin{align}\label{eq:rt}
    V = \frac{(1-\beta){\rm Wi}}{\hat{F}}\int_\mathcal{V}\stackrel{\triangledown}{\dot{\bm{\gamma}}}_0:\bm{\nabla}\hat{\bm{u}}\ d\bm{r} + \mathcal{O}(\rm Wi^2),
\end{align}
where $\hat{\bm{u}}$, $\hat{F}$ are the non-dimensional velocity field and drag of the auxiliary Stokes problem, in which the same rigid particle is translated along $\bm{e}_z$ at unit speed in a Newtonian fluid, and $\dot{\bm{\gamma}}_0$ is the dimensionless strain rate of the Stokes flow. $\mathcal{V}$ represents the fluid volume. 
Equation~\eqref{eq:rt} shows that the propulsion speed is proportional to ${\rm Wi}$ at leading order. The remaining integral is a functional of the particle shape. The central question is therefore which geometric moment of the shape controls this functional at leading order.

Additionally, we introduce a dimensionless functional of the shape alone 
\begin{align}
    \mathcal{J}[r] = \frac{1}{\hat{F}}\int_\mathcal{V} \stackrel{\triangledown}{\dot{\bm{\gamma}}}_0:\bm{\nabla}\hat{\bm{u}}\ d\bm{r},
\end{align}
resulting in $V =(1-\beta) {\rm Wi}\mathcal{J}[r] + \mathcal{O}(\rm Wi^2)$.

For an axisymmetric particle, let $a(z)$ denote the dimensionless cross-sectional radius, so that the particle surface is described by $\rho=a(z)$. The fluid volume element in cylindrical coordinates is $d\bm{r}=\rho d\rho d\theta dz$, and therefore
\begin{align}
    \mathcal{J}[r]= \frac{1}{\hat{F}}\int_0^H\int_0^{2\pi}\int_{a(z)}^\infty\stackrel{\triangledown}{\dot{\bm{\gamma}}}_0:\bm{\nabla}\hat{\bm{u}}\,\rho\,d\rho\,d\theta\,dz .
\end{align}
Since $r^2(z) dz$ defines a probability measure on $[0, H]$, we factor out this measure by defining the effective kernel
\begin{align}
    K(z; [r]) \equiv\frac{1}{\hat{F} r^2(z)}\int_0^{2\pi}\int_{a(z)}^\infty\stackrel{\triangledown}{\dot{\bm{\gamma}}}_0:\bm{\nabla}\hat{\bm{u}}\,\rho\,d\rho\,d\theta,
\end{align}
so that 
\begin{align}
    \mathcal{J}[r]=\int_0^HK(z; [r]) r^2(z) dz.
\end{align}
This recasts $\mathcal{J}$ as a weighted average of $K$ with respect to $r^2(z) dz$, enabling a systematic expansion in moments of the axial volume distribution.

The functional $\mathcal{J}$ must be odd under fore--aft reflection, because a reflected particle propels in the opposite direction, whereas a fore--aft symmetric particle cannot propel. Expanding the shape dependence in central moments of the axial volume distribution, the zeroth moment is fixed by normalization, the first central moment vanishes by definition, and the second moment is even under reflection. The first moment that is both nonzero and odd under reflection is therefore the third central moment. Hence, to leading order in fore--aft asymmetry,
\begin{align}
    \mathcal{J}\sim S + \text{higher-order odd moments}.
\end{align}
Higher-order odd moments contribute subleading corrections. Hence, the leading propulsion velocity is governed by a single geometric scalar, 
\begin{align}
    V \sim {\rm Wi}\cdot S.
    \label{eq:scaling}
\end{align}
Equation~\eqref{eq:scaling} constitutes the central result of this Letter and makes a stronger prediction than the empirical relation in Fig.~\ref{fig3}: the ${\rm Wi}$-dependent slopes should collapse when the velocity is plotted against the combined variable ${\rm Wi}\cdot S$.
Figure~\ref{fig4} confirms this prediction for ${\rm Wi}\leq 1$, where the perturbative theory applies~\cite{Pak2012-rk}. The results for different shapes and Weissenberg numbers collapse onto a single linear master curve, showing that the DNS results are quantitatively organized by the low-${\rm Wi}$ asymptotic theory. This collapse establishes $V\sim{\rm Wi}\cdot S$ as the central scaling law for viscoelastic rotation-induced propulsion.

\textit{Conclusion}---We have shown that rotation-induced propulsion in viscoelastic fluids obeys a universal leading-order scaling $V \sim {\rm Wi}\cdot S$ with $S$ the third central moment of the axial volume distribution. This mechanism relies on normal-stress-induced secondary flows. Because the elastic normal stresses are quadratic in the rotation rate, the propulsion direction is independent of the sign of rotation. As a consequence, reciprocal back-and-forth rotation about a fixed axis---the simplest possible actuation with a single degree of freedom---generates sustained net translation in a viscoelastic fluid. This stands in direct contrast to Purcell's scallop theorem, which forbids net motion under time-reversible actuation in Newtonian fluids, and demonstrates that viscoelasticity enables a fundamentally new propulsion  mode requiring neither chirality nor non-reciprocal actuation.

The propulsion efficiency attainable at optimal conditions, $V/(R\Omega)\approx 10^{-2}$ at $\rm Wi = 1.5$, is comparable to that of biological swimmers such as \textit{E. coli}, which is near optimal in its geometrical design~\cite{Spagnolie2011-ws} and exhibits $V\approx \SI{10}{\micro\meter}\cdot {\rm s}^{-1}$, $R\approx \SI{0.5}{\micro\meter}$ and $\Omega\approx \SI{100}{\hertz}$~\cite{Patteson2015-pa}, yielding $V/(R\Omega)\approx 10^{-2}$. 
Finally, the relation $V\sim{\rm Wi}\cdot S$ suggests an active microrheological application, in which the elastic time scale of the surrounding fluid is inferred from the propulsion speed of a particle with known shape.

\textit{Acknowledgments}---This work was supported by the Grants-in-Aid for Scientific Research (JSPS KAKENHI) under Grant Number 25H01476, by the JSPS Core-to-Core Program `Advanced core-to-core network for the physics of self-organising active matter (JPJSCCA20230002)'. 
T.K. acknowledges the JSPS Overseas Research Fellowship.

\bibliography{ref}


\end{document}